\documentclass[cameraready]{Interspeech}
\title{SRF-SVB: Style-Consistent Singing Voice Beautifying via Rectified Flow}

\author[affiliation={1}, equalcontribution]{Wenhui}{Li}
\author[affiliation={1},equalcontribution]{Biao}{Dong}
\author[affiliation={1},]{Liwei}{Hu}
\author[affiliation={1},]{Jiqing}{Han}
\author[affiliation={1},correspondingauthor]{Yongjun}{He}
\address{
    $^1$ Harbin Institute of Technology, China
}

\email{wenhui.li@stu.hit.edu.cn, heyongjun@hit.edu.cn}


\usepackage{comment}
\usepackage{booktabs}   % 提供 \toprule, \midrule, \bottomrule
\usepackage{multirow}   % 提供 \multirow（跨行）
\usepackage{cite}
\begin{document}

\maketitle

\begin{abstract}
    Singing voice beautifying (SVB) aims to correct pitch and rhythm of amateur singing while enhancing vocal quality, preserving lyrics and the singer's timbre. Existing methods, however, suffer from limited generation quality and efficiency, and tend to neglect the preservation of the singer's style. We propose SRF-SVB, a style-consistent model for SVB via rectified flow, which achieves high-fidelity and efficient beautification covering pitch and rhythm correction. Furthermore, we design a context-guided masked mel-spectrogram inpainting mechanism that effectively preserves the amateur singer's style, including unique timbre and expressive patterns. Experiments on both English and Chinese test sets show that SRF-SVB outperforms baseline models in most objective and subjective metrics.
\end{abstract}

\section{Introduction}

Singing Voice Beautifying (SVB) plays a crucial role in professional music production and online karaoke applications. Traditional approaches rely on professional audio engineers utilizing commercial software such as Autotune\cite{Yong18-Singing} for manual adjustment, a process that is time-consuming, labor-intensive, and requires high technical expertise. Therefore, developing intelligent and automated SVB methods holds significant research value and practical importance. 

In recent years, singing voice research has primarily focused on Singing Voice Conversion (SVC)\cite{Chen24-LDMSVC,Ferreira25-FreeSVC}, which transforms the timbre while preserving melody and lyrics. However, the core objective of SVB is to correct pitch and rhythm of amateur singing while enhancing vocal quality without altering lyrics or the amateur singer's timbre. The task most closely related to SVB is Automatic Pitch Correction (APC)~\cite{Perrotin16-Target,Zhuang22-KaraTuner,Hai23-DiffPitcher}. Representative models in this area, such as KaraTuner\cite{Zhuang22-KaraTuner} and Diff-Pitcher\cite{Hai23-DiffPitcher}, have demonstrated promising results. Nevertheless, they focus solely on pitch while neglecting other crucial aspects of singing quality, such as rhythm and expressiveness. StylePitcher~\cite{Huang25-StylePitcher} proposed a style-following pitch curve generator that preserves original style while generating professional pitch curves through implicit style modeling. NSVB\cite{Liu22-Beauty} was the first work to formally define the SVB task, proposing a generative model that achieves multi-dimensional beautifying based on conditional variational autoencoders (CVAE)\cite{sohn2015learning}. However, it relies on strictly paired singing samples, requiring professional singers to record both amateur and professional versions of the same lyrics, which are extremely difficult to collect in large quantities. Recently, CONTUNER\cite{Wang24-ConTuner} explored diffusion-based SVB without parallel training data, but it struggles to achieve high generation quality while maintaining efficient inference. Rectified flow\cite{Liu22-Rectified,Liu23-Flow} offers a promising alternative by constructing straight ordinary differential equation (ODE) paths from noise to data distribution, significantly improving training stability and inference efficiency compared to diffusion models while maintaining high generation quality. This method has demonstrated superior performance in tasks such as speech synthesis\cite{Guo24-VoiceFlow} and voice conversion\cite{Ren25-ReFlowVC}. Furthermore, the excessive pursuit of professional vocal quality in current SVB methods often weakens the amateur singer's style, including unique timbre and expressive patterns, resulting in a lack of distinctiveness and realism in the beautified singing.

In this paper, we propose a Style-consistent Rectified Flow-based framework for SVB (SRF-SVB). We adopt rectified flow as the generative backbone to ensure high generation quality and efficient inference. To the best of our knowledge, this is the first work introducing rectified flow into the SVB task. To correct pitch and rhythm while preserving lyrics and the singer's timbre, we decouple pitch, timbre, and content features from the singing voice for independent manipulation. Furthermore, to preserve the amateur singer's style, including unique timbre and expressive patterns, we propose a context-guided masked mel-spectrogram inpainting mechanism which randomly masks regions in the mel-spectrogram and reconstructs them using decoupled pitch, timbre, and content features with the surrounding spectral context. This approach ensures style consistency between the generated and surrounding regions. When generating beautified singing, we introduce professional pitch as the target, align the amateur and professional singing temporally via Dynamic Time Warping (DTW)~\cite{Muller07-DTW}, and extract timbre from the amateur singer. The amateur mel-spectrogram serves as context to ensure the generated beautified mel-spectrogram remains style-consistent with the amateur singer.

Our main contributions are summarized as follows:
\begin{figure*}[t]
  \centering
  \includegraphics[width=0.9\textwidth]{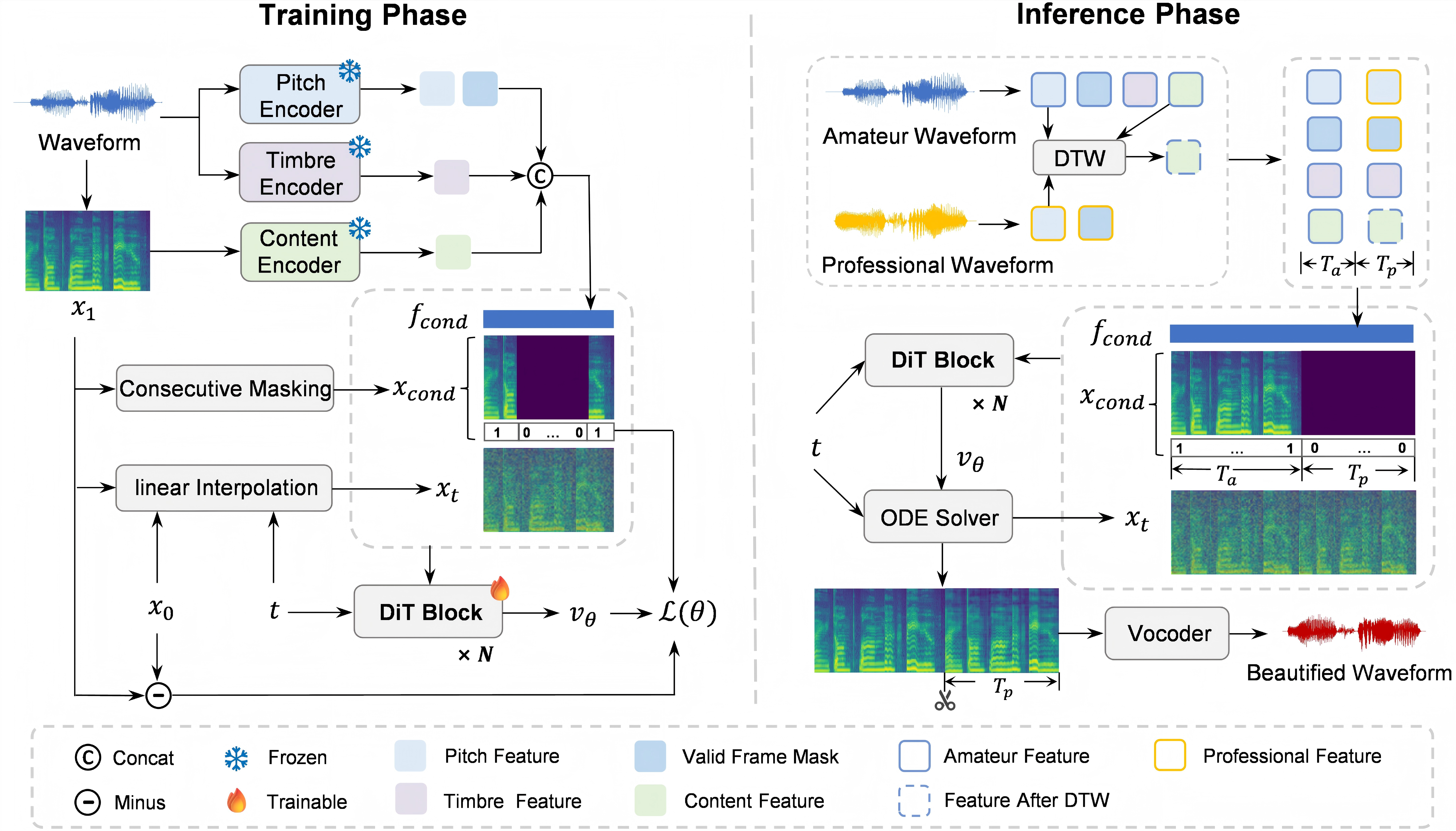}
  \caption{Left: Training procedure of SRF-SVB; Right: Inference procedure of SRF-SVB.}
  \label{fig:architecture}
\end{figure*}

(1) We propose SRF-SVB, the first rectified flow-based SVB framework that achieves high-fidelity and efficient singing voice beautifying, including pitch and rhythm correction.

(2) We design a context-guided masked mel-spectrogram inpainting mechanism to ensure style consistency, enabling natural preservation of the amateur singer's unique timbre and expressive patterns.

(3) Experiments demonstrate that SRF-SVB achieves superior or comparable performance in pitch accuracy, timbre similarity, and overall quality across both English and Chinese test sets in objective and subjective evaluations. Audio samples are available.\footnote{https://mrwho729.github.io/SRF-SVB/}

\section{Proposed Method}

\subsection{Rectified Flow}
Rectified flow\cite{Liu22-Rectified,Liu23-Flow} is a generative modeling framework based on ordinary differential equations. It learns a velocity field to transport samples from a noise distribution ${x}_0 \sim \pi_0$ to the target data distribution ${x}_1 \sim \pi_1$ along straight trajectories. The framework constructs a linear interpolation path:
\begin{equation}
 {x}_t = t {x}_1 + (1 - t) {x}_0
\end{equation}
where $t=0$ corresponds to the pure noise state and $t=1$ corresponds to the target data state.

A neural network is trained to predict the velocity field $v_\theta$ at any time step along this path, satisfying:
\begin{equation}
\frac{d {x}_t}{dt} = v_\theta( {x}_t, t,  {c})
\end{equation}
where $ {c}$ denotes the conditioning information.

The model is optimized by minimizing the following objective function:
\begin{equation}
\mathcal{L}(\theta) = \mathbb{E}_{x_0, x_1, t, c} \left[ \lVert (x_1 - x_0) - v_\theta(x_t, t, c) \rVert_2^2 \right]
\end{equation}
During inference, the ODE is integrated from $t=0$ to $t=1$ to generate target samples. 

\subsection{Model Architecture}
Figure~\ref{fig:architecture} illustrates the overall architecture of SRF-SVB, with training on the left and inference on the right. We first describe the training procedure.

\subsubsection{Acoustic Feature Decoupling}
To control different acoustic attributes independently during beautification, we decouple three types of acoustic features from the input singing voice using dedicated encoders (detailed in Section 3.1.2). The pitch feature ${f}_{p}$ and timbre feature ${f}_{t}$ are extracted from the singing waveform, while the content feature ${f}_{c}$ is extracted from the mel-spectrogram. Additionally, we construct a valid frame mask ${u} \in \{0,1\}^T$ to identify frames with valid fundamental frequency, avoiding interference from silent regions during the generation process. The above four types of features are concatenated along the channel dimension, normalized, and mapped to a unified hidden dimension $H$ through a linear projection layer, yielding the acoustic condition ${f}_{cond} \in \mathbb{R}^{T \times H}$, which guides the subsequent generation process.

\subsubsection{Mel-Spectrogram Masking Strategy}
We obtain the corresponding mel-spectrogram ${x_1} \in \mathbb{R}^{T \times D}$ from the original audio input, where $T$ denotes the number of time frames and $D$ denotes the number of frequency bins. Then we randomly select a starting time position and mask a segment of consecutive frames with length $L = \alpha \cdot T$, where $\alpha$ is a scaling factor controlling the mask ratio. A binary mask vector ${b} \in \{0,1\}^T$ is generated, consistent with the temporal dimension of the mel-spectrogram, where $b_i = 0$ indicates that the $i$-th frame is masked, and $b_i = 1$ indicates that the $i$-th frame is retained as context. This indicator explicitly identifies the missing regions, enhancing the awareness of the model regarding reconstruction locations. Based on this, the masked mel-spectrogram is computed as:
\begin{equation}
{x}_{mask} = {x_1} \odot {b}
\end{equation}
Subsequently, we concatenate ${x}_{mask}$ and ${b}$ along the channel dimension to form the mask-aware condition ${x}_{cond}$.

\subsubsection{Rectified Flow-Based Inpainting}
The velocity field $v_\theta({x}_t, t, {f}_{cond}, {x}_{cond})$ is parameterized based on the Diffusion Transformer (DiT)\cite{Vaswani17-Attention, Peebles23-Scalable} architecture with rotary position embedding\cite{Su24-RoFormer}. Specifically, given time step $t \sim \mathcal{U}(0, 1)$, we first compute the current state ${x}_{t}$ according to the linear interpolation formula between the original mel-spectrogram ${x}_1$ and Gaussian noise ${x}_0 \sim \mathcal{N}(0, I)$. Then we concatenate ${x}_{t}$ with the acoustic condition ${f}_{cond}$ and mask-aware condition ${x}_{cond}$ along the channel dimension to form a joint input representation. The time step $t$ is encoded via an MLP and used to modulate the DiT layers through adaptive layer normalization, enabling fine-grained temporal control over the generation process. The training objective is a conditional flow matching loss:
\begin{equation}
\begin{split}
\mathcal{L}(\theta) = \mathbb{E}_{x_0 \sim \pi_0, x_1 \sim \pi_1, t, f_{cond}} \Big[ \lVert [(x_1 - x_0) \\
- v_\theta(x_t, t, f_{cond}, x_{cond})] \odot (1-b) \rVert_2^2 \Big]
\end{split}
\end{equation}
By computing the loss only over masked regions, the model focuses on learning to inpaint the missing mel-spectrogram content guided by the acoustic condition and mask-aware condition.

\subsection{Inference Procedure}
During inference, the goal is to transform amateur singing into a beautified version while preserving lyrics and the original singer's timbre, as illustrated in the right part of Figure~\ref{fig:architecture}. Given a pair of singing segments with the same content: amateur singing of length $T_a$ and professional singing of length $T_p$, we proceed as follows.

\subsubsection{Acoustic Condition Composition}
The acoustic condition is composed of two parts. For the first $T_a$ frames, which serve as the context region, we use the pitch, timbre, and content features along with the valid frame mask from the amateur singing, ensuring the acoustic condition is consistent with the context mel-spectrogram. For the subsequent $T_p$ frames, which serve as the generation region, professional pitch and valid frame mask are used to guide the model to produce accurate melodic contours; content features are mapped to the professional time axis using the alignment obtained by DTW between the amateur and professional pitch curves to achieve rhythmic correction; and timbre features from the amateur singer are retained to ensure timbre consistency. Finally, features and masks from both segments are concatenated along the temporal dimension to obtain the complete acoustic condition $f_{cond} \in \mathbb{R}^{({T_a}+{T_p}) \times H}$ for inference.

\subsubsection{Mask-Aware Condition Construction}
We construct a masked mel-spectrogram with a total length of ${T_a}+{T_p}$, denoted as $x_{mask} \in \mathbb{R}^{({T_a}+{T_p}) \times D}$, where the first $T_a$ frames contain the mel-spectrogram of the amateur singing serving as context, while the subsequent $T_p$ frames are initialized to silence, representing the region to be generated.  A binary mask vector $b \in \{0,1\}^{T_{a}+T_{p}}$ is constructed accordingly, where the first $T_a$ elements are set to 1 and the last $T_p$ elements are set to 0, indicating that the model should focus on the reconstruction in the latter region. The mask-aware condition $x_{cond}$ is formed by concatenating $x_{mask}$ and $b$ along the channel dimension.

\subsubsection{Beautified Singing Generation}
Starting from Gaussian noise with the same shape as $x_{mask}$ at $t=0$, the state is iteratively updated by ODE solver using the velocity field predicted by the DiT network based on the current state, guided by the acoustic condition and mask-aware condition, until $t=1$. Finally, we obtain the complete output mel-spectrogram and extract only the last $T_p$ frames as the beautified mel-spectrogram, which is then synthesized into the beautified singing waveform through a pre-trained neural vocoder.

\section{Experiments}
\subsection{Experimental Setup}
\subsubsection{Datasets}
We employ multiple publicly available Chinese and English singing datasets for training, including PopBuTFy\cite{Liu22-Beauty}, GTSinger\cite{Zhang24-GTSinger}, OpenSinger\cite{Huang21-MultiSinger}, Opencpop\cite{Wang22-Opencpop}, PopCS\cite{Liu22-DiffSinger}, and M4Singer\cite{Zhang22-M4Singer}, totaling approximately 160 hours of high-quality singing segments. 

For evaluation, our test data comprises paired amateur and professional singing segments. We adopt the English test set containing 617 paired segments along with their lyric labels from PopBuTFy\cite{Liu22-Beauty}. Additionally, we construct a Chinese test set containing 874 paired segments by: (1) pairing amateur singing segments from CCMusic\cite{Zhou25-CCMusic} with dry vocals extracted from professional songs using Ultimate Vocal Remover~\cite{UVR5}; (2) recording amateur versions ourselves to pair with selected professional singing segments from Opencpop~\cite{Wang22-Opencpop} and OpenSinger~\cite{Huang21-MultiSinger}. MIDI information is extracted by SOME\cite{OpenVPI22-SOME} from professional segments for pitch accuracy evaluation, and lyric labels are manually generated for content evaluation. All test samples are excluded from the training set.

\subsubsection{Implementation Details}
We train the model on a single NVIDIA GeForce RTX 4090 GPU with 24GB memory for 500k steps, employing dynamic batch size allocation to fully utilize GPU resources. We use the AdamW\cite{Loshchilov19-Decoupled} optimizer. For the learning rate schedule, the first 10k steps perform linear warm-up from 0 to $2.5 \times 10^{-4}$, followed by exponential decay to $1 \times 10^{-4}$. The mask ratio $\alpha$ is set to 0.5.

For acoustic feature extraction, we employ RMVPE\cite{Wei23-RMVPE} as the pitch extractor, use a Conformer\cite{Gulati20-Conformer} model pre-trained on singing voice datasets to extract PPG (Phonetic Posterior Gram)\cite{Sun16-Phonetic} as content features, and adopt the speaker verification model CAM++\cite{Wang23-CAM} to extract timbre embeddings. All audio is processed at a 22050 Hz sampling rate with a frame length of 1024, hop size of 128, and 80 mel bins. For the vocoder, we adopt NSF-HiFiGAN\cite{Kong20-HiFiGAN} to convert mel-spectrograms into audio waveforms.

For network architecture, the DiT backbone consists of 12 Transformer blocks with a hidden dimension of 512 and 8 attention heads per layer. During inference, we use Euler's method to solve the ODE with 10 sampling steps.

\subsubsection{Baseline Models}
We select two representative baseline models for comparison: (1) Diff-Pitcher\cite{Hai23-DiffPitcher}, a diffusion-based automatic pitch correction model that focuses on pitch dimension correction; (2) NSVB\cite{Liu22-Beauty}, a generative SVB model based on CVAE, which requires parallel amateur-professional training data. 

\begin{table*}[t]
  \centering
  \caption{Objective and subjective evaluation results of different models on English and Chinese test sets.}
  \label{tab:results}
  \resizebox{\linewidth}{!}{
  \begin{tabular}{lcccccccccc}
    \toprule
    \multirow{2}{*}{\textbf{Model}} & \multicolumn{5}{c}{\textbf{English}} & \multicolumn{5}{c}{\textbf{Chinese}} \\
    \cmidrule(lr){2-6} \cmidrule(lr){7-11}
    & RPA$\uparrow$ & CER$\downarrow$ & SECS$\uparrow$ & MOS-Q$\uparrow$ & MOS-S$\uparrow$ & RPA$\uparrow$ & CER$\downarrow$ & SECS$\uparrow$ & MOS-Q$\uparrow$ & MOS-S$\uparrow$ \\
    \midrule
    Amateur & 0.40 & 0.22 & - & 3.60$\pm$0.08 & - & 0.22 & \textbf{0.02} & - & 3.42$\pm$0.08 & - \\
    Diff-Pitcher & 0.48 & \textbf{0.20} & 0.68 & 2.45$\pm$0.16 & 3.44$\pm$0.22 & 0.37 & 0.03 & 0.80 & 1.96$\pm$0.13 & 2.72$\pm$0.24 \\
    NSVB & 0.57 & 0.23 & 0.58 & 3.76$\pm$0.17 & 3.83$\pm$0.12 & 0.47 & 0.05 & 0.40 & \textbf{3.64$\pm$0.20} & 3.18$\pm$0.12 \\
    SRF-SVB (ours) & \textbf{0.57} & 0.24 & \textbf{0.85} & \textbf{3.91$\pm$0.13} & \textbf{4.47$\pm$0.17} & \textbf{0.50} & 0.04 & \textbf{0.82} & 3.62$\pm$0.19 & \textbf{4.06$\pm$0.22} \\
    \bottomrule
  \end{tabular}
  }
\end{table*}
\subsubsection{Evaluation Metrics}
Objective metrics cover three dimensions: (1) Raw Pitch Accuracy (RPA): the percentage of frames where pitch deviation from ground-truth MIDI falls within half a semitone; (2) Character Error Rate (CER): computed by comparing FireRedASR\cite{Xu25-FireRedASR} recognition results against lyric labels to evaluate content preservation; (3) Speaker Embedding Cosine Similarity (SECS): cosine similarity between CAM++ embeddings of beautified and amateur audio to evaluate timbre consistency.

For subjective evaluation, 15 volunteers rate 30 samples (evenly split between English and Chinese) covering various singers on a 5-point Mean Opinion Score (MOS) scale across two dimensions: (1) Quality MOS (MOS-Q): evaluating audio clarity, naturalness, and overall sound quality; (2) Similarity MOS (MOS-S): measuring whether the beautified singing retains the original singer's timbre and expressive patterns. All MOS results are reported with 95\% confidence intervals. For the ablation study, we adopt Comparative Mean Opinion Score (CMOS) ranging from -3 to 3, where listeners directly compare system variants against the full model on both quality (CMOS-Q) and similarity (CMOS-S) dimensions.

\begin{table}[t]
  \centering
  \caption{Ablation study on masking strategies.}
  \label{tab:ablation}
  \resizebox{\columnwidth}{!}{
  \begin{tabular}{lccc}
    \toprule
    \textbf{Masking Strategy} & SECS$\uparrow$ & CMOS-Q & CMOS-S \\
    \midrule
    \multicolumn{4}{l}{\textbf{English}} \\ 
    \hspace{0.2cm}Consecutive ($\alpha$=0.5, ours) & \textbf{0.85} & \textbf{0.00} & \textbf{0.00} \\
    \hspace{0.2cm}Consecutive ($\alpha$=0.2) & 0.77 & -0.53 & -0.27 \\
    \hspace{0.2cm}Random & 0.76 & -0.87 & -0.50 \\
    \midrule
    \multicolumn{4}{l}{\textbf{Chinese}} \\
    \hspace{0.2cm}Consecutive ($\alpha$=0.5, ours) & \textbf{0.82} & \textbf{0.00} & \textbf{0.00} \\
    \hspace{0.2cm}Consecutive ($\alpha$=0.2) & 0.70 & -0.67 & -0.60 \\
    \hspace{0.2cm}Random & 0.71 & -0.97 & -1.03 \\
    \bottomrule
  \end{tabular}
  }
\end{table}

\subsection{Results}
Table \ref{tab:results} presents the objective and subjective evaluation results on both English and Chinese test sets. Regarding pitch accuracy, the amateur singing exhibits low RPA scores (0.40 for English and 0.22 for Chinese), reflecting the prevalent pitch deviation issues. After beautification, SRF-SVB achieves the highest RPA scores on both test sets, reaching 0.57 for English and 0.50 for Chinese, outperforming 
Diff-Pitcher and NSVB by 35.1\% and 6.4\% on Chinese respectively. It demonstrates that our method more effectively corrects amateur pitch contours toward professional-level accuracy.

In terms of content preservation, SRF-SVB yields a slightly higher CER than Diff-Pitcher on the English test set, primarily because Diff-Pitcher performs pitch correction only, whereas our method employs a comprehensive generative reconstruction framework. On the Chinese test set, all methods achieve comparably low CER values, indicating satisfactory content preservation across different models.

Most notably, SRF-SVB outperforms all baseline models in timbre similarity, achieving SECS scores of 0.85 and 0.82 on the English and Chinese test sets, respectively. In contrast, NSVB exhibits substantially lower timbre similarity (0.58 for English and 0.40 for Chinese), confirming that it sacrifices the singer's unique timbre characteristics while pursuing professional-level singing quality. Our proposed context-guided masked mel-spectrogram inpainting mechanism successfully preserves the amateur singer's individual timbre by leveraging spectral context information.

For subjective evaluation, SRF-SVB obtains the highest MOS-Q on English 
(3.91) and comparable performance on Chinese (3.62, compared to 3.64 
for NSVB). More importantly, SRF-SVB significantly outperforms all 
baselines in MOS-S, demonstrating superior preservation 
of the singer's timbre and expressive patterns. Diff-Pitcher receives 
the lowest scores across all metrics, indicating that pitch-only 
correction is insufficient for high-quality beautification.

\subsection{Ablation Study}

Table \ref{tab:ablation} presents the ablation study results on masking strategies. We compare our consecutive masking ($\alpha$=0.5) against two variants: consecutive masking with a smaller ratio ($\alpha$=0.2), and random masking that masks 50 non-contiguous segments while maintaining the same total mask ratio of 0.5.

Results show that our masking strategy consistently outperforms both variants, achieving SECS scores of 0.85 and 0.82 on English and Chinese, compared to 0.77 and 0.70 for consecutive masking ($\alpha$=0.2), and 0.76 and 0.71 for random masking. The CMOS evaluation confirms this trend, with random masking showing the most severe degradation, reaching CMOS-S of -1.03 on Chinese. The consecutive masking ($\alpha$=0.2) limits the generation scope, while random masking fragments context into short segments, preventing the model from learning long-range dependencies. Moreover, our masking strategy best ensures training-inference consistency, enabling effective timbre retention and superior generation quality.

\section{Conclusion}
In this paper, we presented SRF-SVB, the first style-consistent rectified flow-based model for efficient and high-fidelity singing voice beautification. We designed a context-guided masked mel-spectrogram inpainting mechanism that effectively preserves the amateur singer's style, including timbre and expressive patterns. Experiments on both English and Chinese datasets demonstrate that SRF-SVB achieves superior or comparable performance in pitch accuracy, timbre similarity, and overall quality. However, we observe that the generative process may occasionally impact pronunciation clarity, as indicated by the CER results. Future work will focus on enhancing linguistic fidelity and optimizing inference efficiency for real-time applications such as live streaming and online karaoke.

\section{Generative AI Use Disclosure}
Generative AI tools were used to assist with grammar checking, 
language polishing, and editing of this manuscript. All scientific 
content, including research ideas, methodology, experiments, and analysis, 
was solely produced by the authors. The authors take full responsibility 
for the content of this paper.

\bibliographystyle{IEEEtran}
\bibliography{mybib}

\end{document}